\documentclass[conference, a4paper, 10pt]{IEEEtran}

\usepackage[british,UKenglish]{babel}       
\usepackage[utf8]{inputenc}
\usepackage{csquotes}
\usepackage[T1]{fontenc}
\usepackage{textcomp}
\usepackage{microtype}
\usepackage{ragged2e}
\usepackage{mathtools}                      
\usepackage{amssymb}
\usepackage{amsfonts}
\usepackage[cmintegrals]{newtxmath}
\usepackage{ntheorem}
\usepackage{algorithm}
\usepackage{algorithmic}
\usepackage{comment}
\usepackage{cite}
\usepackage{fancyref}
\usepackage{balance}
\usepackage{enumitem}                       
\usepackage[
    acronym,
    toc,
    nopostdot,
    nogroupskip,
    nonumberlist,
    ]{glossaries}
\glsdisablehyper
\newacronym{adc}{ADC}{analog-to-digital converter}

\newacronym{dac}{DAC}{digital-to-analog converter}

\newacronym{lut}{LUT}{look-up table}

\newacronym{rf}{RF}{radio frequency}

\newacronym{ue}{UE}{user equipment}

\newacronym[shortplural = AAS, firstplural = advanced antenna systems (AAS)]{aas}{AAS}{advanced antenna system}

\newacronym{dpd}{DPD}{digital pre-distortion}

\newacronym{pimc}{PIMC}{passive intermodulation correction}

\newacronym[shortplural = MIMO, firstplural = multiple inputs multiple outputs (MIMO)]{mimo}{MIMO}{multiple input multiple output}

\newacronym{csi}{CSI}{channel state information}

\newacronym{bpa}{BPA}{beam pattern alignment}

\newacronym{rmse}{RMSE}{root-mean-square error}

\newacronym[shortplural = VNA, firstplural = vector network analysers (VNAs)]{vna}{VNA}{vector network analyser}

\newacronym{bfw}{BFW}{beamforming weights}

\newacronym{dbf}{DBF}{fully-digital beamforming}

\newacronym{abf}{ABF}{analog beamforming}

\newacronym{hbf}{HBF}{hybrid beamforming}

\newacronym{dl}{DL}{downlink}

\newacronym{ul}{UL}{uplink}

\newacronym{tdd}{TDD}{time division duplex}

\newacronym{fdd}{FDD}{frequency division duplex}

\newacronym[shortplural = RIS, firstplural = reconfigurable intelligent surfaces (RIS)]{ris}{RIS}{reconfigurable intelligent surface}

\newacronym[shortplural = AOA, firstplural = angles of arrival (AOA)]{aoa}{AOA}{angle of arrival}

\newacronym{pca}{PCA}{principal component analysis}

\newacronym[shortplural = TXs, firstplural = transmitters (TXs) ]{tx}{TX}{transmitter}

\newacronym[shortplural = RXs, firstplural = receivers (RXs) ]{rx}{RX}{receiver}

\newacronym{ota}{OTA}{over-the-air}

\newacronym[shortplural = DNN, firstplural = deep neural networks (DNN)]{dnn}{DNN}{deep neural network}

\newacronym[shortplural = GPs, firstplural = Gaussian processes (GPs)]{gp}{GP}{Gaussian process}

\newacronym[shortplural = APs, firstplural = access points (APs)]{ap}{AP}{access point}

\newacronym{los}{LOS}{line-of-sight}

\newacronym{nlos}{NLOS}{Non-line-of-sight}

\newacronym{ard}{ARD}{Automatic Relevance Determination}

\newacronym{nlml}{NLML}{Negative Log Marginal Likelihood}

\newacronym{rkhs}{RKHS}{reproducing kernel Hilbert space}

\newacronym{rbf}{RBF}{radial basis function}

\newacronym{se}{SE}{squared exponential}

\newacronym{rq}{RQ}{rational quadratic}

\newacronym{sm}{SM}{spectral mixture}

\newacronym{gmm}{GMM}{Gaussian mixture model}

\usepackage{array}                          
\usepackage{booktabs}
\usepackage{longtable}
\usepackage{tabularx}  
\usepackage{threeparttablex}
\usepackage{multirow}
\usepackage[font=small]{caption}            
\usepackage{float}
\usepackage{placeins}
\usepackage{subcaption}
\usepackage{rotating}
\usepackage{stfloats}
\usepackage{hyperref}                       
\hypersetup{
    colorlinks=true,
    linkcolor=blue,
    citecolor=red,
    urlcolor=magenta,
}
\usepackage{url}
\usepackage{bookmark}
\usepackage{graphicx}                       
\graphicspath{{./figures/}}
\DeclareGraphicsExtensions{.pdf,.png,.jpeg,.jpg,.eps,.svg}
\usepackage[dvipsnames]{xcolor}
\usepackage{tikz}                           
\usepackage{pgfplots}
\usepackage{tikz-cd}
\usepackage{bloques}
\usepackage{datatool}                       
\usepackage{textmerg}
\usepackage{gitinfo2}                       

\newcommand{\etal}{et al. }                 

\begin{document}

\title{
    Antenna Array Calibration Via \\ Gaussian Process Models
}

\author{
    Sergey~S.~Tambovskiy$^{\dag\ddag}$,
    Gábor~Fodor~$^{\dag\ddag}$,
    Hugo~M.~Tullberg$^{\dag}$\\
    \small $^\dag$Ericsson Research, Stockholm, Sweden. E-mail: \texttt{Sergey.Tambovskiy|Gabor.Fodor|Hugo.Tullberg@ericsson.com}\\
    \small $^\ddag$KTH Royal Institute of Technology, Stockholm, Sweden. E-mail: \texttt{sergeyta|gaborf@kth.se}
}

\maketitle

\begin{abstract}

    Antenna array calibration is necessary to maintain the high fidelity of beam patterns across a wide range of advanced antenna systems and to ensure channel reciprocity in time division duplexing schemes.
    Despite the continuous development in this area, most existing solutions are optimised for specific radio architectures, require standardised over-the-air data transmission, or serve as extensions of conventional methods.
    The diversity of communication protocols and hardware creates a problematic case, since this diversity requires to design or update the calibration procedures for each new advanced antenna system.
    In this study, we formulate antenna calibration in an alternative way, namely as a task of functional approximation, and address it via Bayesian machine learning.
    Our contributions are three-fold.
    Firstly, we define a parameter space, based on near-field measurements, that captures the underlying hardware impairments corresponding to each radiating element, their positional offsets, as well as the mutual coupling effects between antenna elements.
    Secondly, Gaussian process regression is used to form models from a sparse set of the aforementioned near-field data. 
    Once deployed, the learned non-parametric models effectively serve to continuously transform the beamforming weights of the system, resulting in corrected beam patterns.
    Lastly, we demonstrate the viability of the described methodology for both digital and analog beamforming antenna arrays of different scales and discuss its further extension to support real-time operation with dynamic hardware impairments.
  
\end{abstract}


\section{Introduction}\label{INTRO}
Successfully deploying novel \glspl{aas} requires the
correction of \gls{rf} hardware impairments originating from various sources.
Among those are phase noise, active and passive intermodulation distortions, in-phase/quadrature imbalances and manufacturing imperfections of antenna arrays \cite{EfficientChannelEstimation_Wu.etal_2019, AmplitudePhaseEstimation_Tian.etal_2020, ImpactCalibrationNonlinear_Nie.etal_2021}.
In the latter case, \gls{rf} connections between \gls{adc}/\gls{dac} units and radiating elements can have unmatched lengths and other defects, resulting in different multiplicative complex gain and phase distortions across the signal bandwidth (e.g. group delays, phase delays).
Furthermore, erroneous antenna element placement causes mutual coupling effects.
Consequently, in \gls{tdd} systems, the \gls{dl} and \gls{ul} wireless channel responses, measured within the same coherence interval, become different, entailing that the reciprocity property no longer holds and the wireless channel estimation procedure breaks \cite{MassiveMIMOSelfCalibration_Luo.etal_2019}.

Without a doubt, these hardware defects also negatively affect \gls{fdd}, \gls{abf}, \gls{hbf}, reconfigurable intelligent surfaces and other systems.
Notably, the general fidelity of beamforming patterns \cite{CalibrationPhaseShifter_Wei.etal_2020, IntermodulationDistortionOriented_Aoki.etal_2022, NeuralNetworkBasedPhase_Iye.etal_2022, KnowYourChannel_Moon.etal_2020} and the quality of source detection with angle of arrival estimation algorithms \cite{AmplitudePhaseEstimation_Tian.etal_2020} are negatively affected.
The compensation of these specific distortions falls into the domain of antenna array calibration models and algorithms.

\subsection*{\textbf{\textit{Related works and historical development.}}}

The concepts defining array calibration first appeared in the context of military radar arrays.
Among those are the mutual coupling and blind methods.
For example, Willerton and Manikas \cite{Arrayshapecalibration_Willerton.Manikas_2011} improved the concept of blind array calibration for multi-carrier cases by reducing the required amount of external \gls{ul} signals to a single moving \gls{ue}.
This correction scheme takes into account all of the \gls{rf} impairments of the array and requires no additional form of signal overhead.
While this operating principle makes it applicable in both \gls{tdd} and \gls{fdd} systems, it lacks support of transmitter calibration, and its computational complexity significantly increases with the number of receivers.

The mutual coupling method, also known as self-calibration, is represented by a large volume of recent works in the field. 
Luo \etal \cite{MassiveMIMOSelfCalibration_Luo.etal_2019} provide a comprehensive analysis of this methodology.
First, the authors use combinatorial optimisation to find 
an interconnection network of mutual coupling measurements. 
Then, they prove that the ``star'' interconnection topology between radiating elements is optimal for fully calibrating \gls{aas} with arbitrary geometry.
Wang \etal \cite{ArrayErrorsAntenna_Wang.etal_2021} address the common disadvantage of mutual coupling calibration by jointly estimating array errors and antenna element patterns.

\Gls{ota} calibration is another family of methods that rely exclusively on external feedback information from the \gls{ue}.
As such, Wei \etal \cite{CalibrationPhaseShifter_Wei.etal_2020} exploit the \gls{ota} concept by relying on \gls{ul} pseudo-noise training sequences in a line-of-sight setting to calibrate a network of \gls{abf} phase shifters.
Their solution allows for the performance estimation of the respective phase deviations with affordable computational complexity, and the proposed algorithms include Cramer-Rao lower bound estimates for a number of required measurements.
In a tangential work, Tian \etal \cite{AmplitudePhaseEstimation_Tian.etal_2020} 
study statistical properties of \gls{ota} estimators for both amplitude scaling and phase drifts within the \gls{rf} chains.
Lastly, Moon \etal \cite{KnowYourChannel_Moon.etal_2020} address the disadvantages of other \gls{ota} solutions by leveraging existing wireless channel estimation protocols.
In that work, relative phase mismatches are calibrated by performing the channel estimation with \gls{ue} assistance while continuously adjusting phase values across the array.

With the surge in data-driven modelling, \glspl{dnn} have become applicable in the domain of array calibration. 
Shan \etal \cite{DiagnosisCalibrationState_Shan.etal_2019} apply the encoder-decoder \gls{dnn} to the \gls{dl} pilot matrix to reconstruct measured \gls{rf} impairments. 
Despite experimental accuracy, its training procedure relies on the assumption that the measured \gls{rf} impairment dataset will not be affected by distribution shift.
Thus, the capability of the aforementioned calibration scheme in an online setting remains open.
Furthermore, it is still undetermined how well that algorithm scales with an increasing number of antenna elements and the complexity of the precoder.
In most recent work, Iye \etal \cite{NeuralNetworkBasedPhase_Iye.etal_2022} propose a fully-connected \gls{dnn}, which utilises a large dataset consisting of measured beam patterns.
Once the computationally extensive \gls{dnn} is trained, the phase errors in the array are estimated using only a single beam pattern as an input.
The latter indicates that, despite being limited only to phase estimation, the method is more time-efficient than conventional calibration procedures.

\subsection*{\textbf{\textit{Contributions.}}}

Our contributions start with a definition of the measured parameter space that captures the underlying \gls{rf} hardware impairments corresponding to each radiating element, their positional offsets, as well as the mutual coupling effects between them.
Next, we use \gls{gp} regression on a sparse set of data to approximate the surfaces of near-field measurements.
Afterwards, learned non-parametric models are deployed to continuously transform \gls{bfw} of the system, resulting in corrected beam patterns and calibrated \gls{aas}.
Lastly, we demonstrate the viability of the described methodology for both \gls{abf} and \gls{dbf} antenna arrays of different scales and discuss its further extension to support operation with dynamic hardware impairments.

\section{Requirements, Limitations and \\ Performance Metrics}
\label{CRPM}
\subsection*{\textbf{\textit{Requirements and Limitations.}}}

Conventional calibration models and techniques are based around \gls{rf} hardware design decisions -- such as mutual coupling properties in symmetrical antenna arrays -- or are designed to operate in the context of specialised communication protocols between \glspl{ap} and \gls{ue}.
The aforementioned paradigms, combined with the increasing diversity of 6G communication protocols and hardware, create a problematic scenario, which necessitates to design new calibration procedures or to update the existing  procedures for each new \gls{aas}.
We postulate that in order to devise a calibration solution for \glspl{aas} with good generalisation capabilities, a set of technical requirements, which future algorithms should support, must be established.

First, the environment in which \glspl{aas} typically operate is likely to have varying temperature conditions, resulting in stochastic changes in the parameters of the \gls{rf} circuitry.
Furthermore, \gls{rf} hardware may degrade with time, causing similar effects.
The presence of both conditions implies that the methodology must support these two types of dynamics.

Calibration belongs to the family of hardware impairment correction algorithms that includes \gls{dpd}, \gls{pimc} and phase noise compensation.
In rare cases, they are combined to facilitate each other's performance, but given the variety of models, hybrid methods can hardly lead to a universally applicable algorithm.
Therefore, calibration must operate independently from the above-mentioned algorithms.
The same reasoning can be applied to the combinations of precoding schemes (beamforming) and \gls{aas} calibration. 

Solutions which rely on a feedback from \gls{ue} or on a standardised data exchange can converge and maintain an ``absolute'' (unbiased) calibration on both ends of the communication link. 
Despite their advantage, such an approach may impose constraints on the protocol design, leading to increased latency and standardisation challenges.
Therefore, a dependency on any type of feedback from the \gls{ue} has to be avoided.

Massive \gls{mimo} in the context of 6G systems have the potential to support antenna arrays with non-uniform (asymmetrical) geometries with thousands of radiating elements.
Additionally, \gls{rf} front-ends are structured to support either \gls{dbf}, \gls{abf} or \gls{hbf} schemes.
It immediately leads to three limitations of conventional antenna calibration: (1) unfeasible large-scale measurement procedures, (2) inaccessibility to mutual coupling and (3) scalability with growing array dimensions.
Consequently, a universal and practically useful calibration must utilise a scalable modular framework to meet the aforementioned limitations.

\subsection*{\textbf{\textit{Performance Metrics.}}}

The quality of array calibration procedures can be assessed in multiple ways.
We propose to classify calibration procedures into two categories: indirect and direct methods.
The first, for example, requires an estimation of the improvement in terms of spectral efficiency. 
As such, indirect methods require complex simulators,
where a multitude of other parameters can affect the results, which,
if put in the context of field trials, require \gls{ue} transceivers 
with running baseband processing.

The second category includes feedback-based measurements of complex-valued frequency responses for all antenna channels.
In practice, feedback \gls{rf} chains connect with the respective antennas before the radiating elements.
Thus, mutual coupling and element offsets are not taken into account during the measurement procedures.
Internal \gls{rf} feedbacks have been utilised for conventional \gls{mimo}, but are no longer viable in practice due to high hardware complexity and the scale of massive \gls{mimo} systems.

In simulations, it is readily possible to access all metrics and controllable variables, but \gls{aas} calibration should still be verifiable, regardless of the chosen methodology or internal \gls{rf} design.
Considering the constraints and requirements discussed above, the most suitable remaining option is to measure 
beam patterns before and after the applied calibration, and subsequently compare them with the desired beam shape.
Based on the above considerations, we define the \gls{bpa} (\ref{bpa-rmse}) as the \gls{rmse} 
computed between third-order tensors of the desired (ideal) $P \in \mathbb{R}^{I \times J \times K}$ and measured $\hat{P}$ beam patterns as:

\begin{equation}
    \mathrm{{\textit{BPA}}_{\mathrm{RMSE}}}=\sqrt{\frac{\sum_{i=1}^{I}\sum_{j=1}^{J}\sum_{k=1}^{K}\left(P_{i,j,k}-\hat{P}_{i,j,k}\right)^{2}}{I+J+K}},
    \label{bpa-rmse}    
\end{equation}

where $I, J, K$ denote azimuth, elevation and frequency.
From a practical perspective, this metric only requires a single scanning probe 
and has a relatively straightforward measurement procedure.
It is worth noting that across recent academic studies, 
there has been a noticeable growth in works that prioritise beam pattern evaluations 
over other performance indicators. 

\section{System Model}\label{SYSMOD}
\begin{figure}[]
    \centering
    \captionsetup{justification=centering}
    \includegraphics[scale=1.0, keepaspectratio]{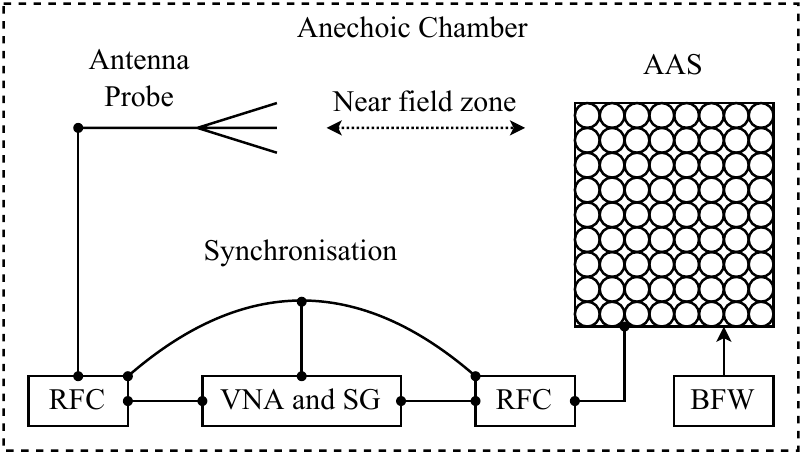}
    \caption{$S_{12}/S_{21}$ -- parameter measurement setup.}
    \label{FIG_system}
\end{figure}

In this paper we consider a system setup that allows to measure complex-valued scattering $S_{12}$ and $S_{21}$ parameters between the probe antenna and the \gls{aas} array.
Fig. \ref{FIG_system} depicts our system that contains the following components: a movable antenna probe (e.g. horn antenna) that can be placed in the near field zone of each isotropic radiating element of the antenna array, a \gls{vna}, a signal generator (SG), a synchronised \gls{rf} circuitry (RFC) chosen according to operating frequency specifications, and a processing unit that assigns \gls{bfw} to the elements of the array.

\begin{figure}[]
    \centering
    \captionsetup{justification=centering}
    \includegraphics[scale=1.0, keepaspectratio]{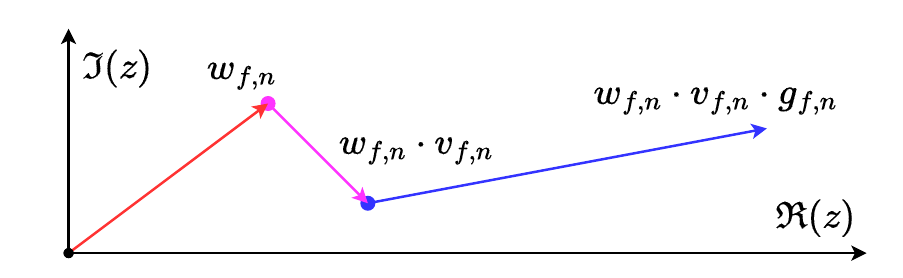}
    \caption{Distortion of a single beamforming weight into a non-calibrated state. 
    Only the resulting vectors are shown.}
    \label{FIG_cmpx_dist}
\end{figure}

In the sequel, we assume that measurements are made within an anechoic chamber without external interference sources and that the internal non-linear distortions (intermodulation components) are corrected by their respective algorithms (e.g. \gls{dpd}, \gls{pimc}).
Therefore, the resulting beam pattern is only affected by frequency-selective linear multiplicative effects applied at each \gls{rf} channel feeding to its antenna -- $z_{f}$ and imperfections of an array itself -- $g_{f}$.
In practice, such distortions depend on the temperature and dynamics caused by hardware degradations, 
which in this study are assumed to be static (memoryless).
The effects of $v_{f}$ and $g_{f}$ on the \gls{bfw} are shown in Fig. \ref{FIG_cmpx_dist}, and are defined as:

\begin{equation}
    w^{d}_{f,n} = w_{f,n} \cdot v_{f,n} \cdot g_{f,n}; \ \ \ w, w^d, v, g \in \mathbb{C}, 
    \label{eq_clb_dist}
\end{equation}

where $f$ denotes the operating frequency point or a sub-carrier within a bandwidth, 
$n$ denotes antenna number, while $w_{f,n}$ and $w^{d}_{f,n}$ denote the desired and distorted \gls{bfw} respectively.
Note that the set of $w^{d}_{f,n}$ values is obtained directly from the measured $S_{12}$ or $S_{21}$ parameters.
The \Gls{aas} under test supports two beamforming modes: \gls{dbf} and emulated \gls{abf}.
In the latter case, each antenna element is assigned adjustable discrete phase and amplitude values, which are ``unique'' for the whole operating bandwidth.

\begin{figure*}[]
    \centering
    \includegraphics[scale=0.27]{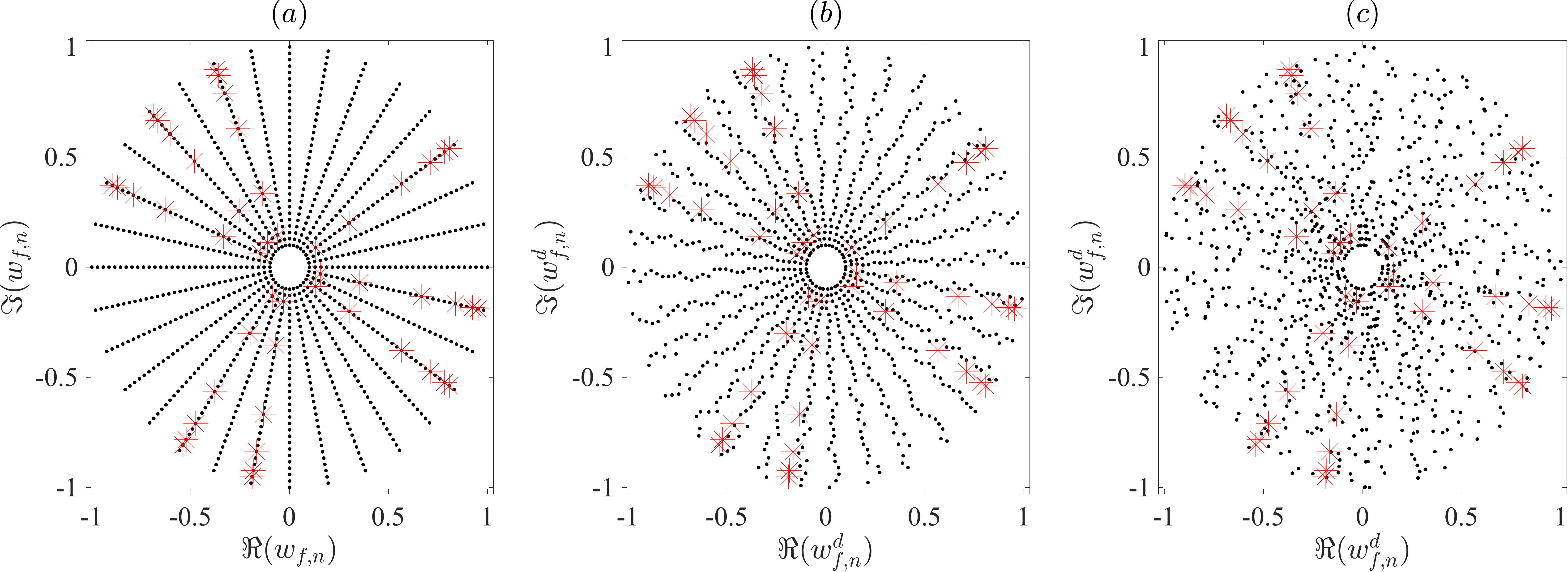}
    \captionsetup{justification=centering}
    \caption{(a) -- $w_{f,n}$ unaffected by distortion. (b) -- small-scale distortion. (c) large-scale distortion. Black dots -- available \gls{abf} weights. Red stars -- subset chosen for \gls{abf}. Each channel has its own distortion pattern.}
    \label{FIG_bf_space_Tri}
\end{figure*}

The effects of distortion can be further visualised in the following example.
Consider an  \gls{abf} system with 64 radiating elements and $5$-bit precision, which allows for 32 gain values and 32 phase states resulting in 1024 complex-valued $w_{f, n}$, to choose from for each antenna.
As shown in Fig. \ref{FIG_bf_space_Tri}\textcolor{blue}{a}, $w_{f, n}$ (red stars) can be allocated across available domain (black dots).
Although discretisation in \gls{abf} negatively affects the radiation patterns, an appropriate increase in the precision range allows to reach the desired \gls{dbf} quality.
Once the studied distortion is induced, as shown in Fig. \ref{FIG_bf_space_Tri}\textcolor{blue}{b} and Fig. \ref{FIG_bf_space_Tri}\textcolor{blue}{c}, the resulting $w^{d}_{f,n}$ weights no longer correspond to the \gls{bfw} that 
were assumed during the beamforming synthesis of $w_{f,n}$.

This observation gives rise to our \textit{\textcolor{red}{first question}}: \textit{how to perform \gls{aas} calibration, while satisfying an extensive set of technical constraints?}

\section{Calibration as High-Dimensional Regression}\label{PROFORM}
To answer the first question, we have to look at calibration with a scanning probe \cite{Calibrationmethodsphased_Seker_2013}.
By design, it is used to create a set of initial conditions for mutual coupling and other calibration methods.
The probe connected to the measurement hardware is placed in front of an active antenna 
element of the \gls{aas}, while others are in inactive states.
Scattering parameters are measured, and the estimated frequency responses are stored in a static \gls{lut}.
These measurements reliably capture all necessary information to model the behaviour of \gls{rf} imperfections for the purposes of array calibration.
However, the scale of massive \gls{mimo} systems, the number of discrete \gls{bfw} in \gls{abf} cases, combined with complex test setup and unsupported continuous calibration, make such a direct approach unfeasible.
This leads to our \textit{\textcolor{red}{second question}}: \textit{how to build upon ``\gls{lut}'' calibration to satisfy modern constraints?}

\subsection*{\textbf{\textit{{Problem Formulation.}}}}

\begin{figure}[]
    \centering
    \captionsetup{justification=centering}
    \includegraphics[scale=0.9, keepaspectratio]{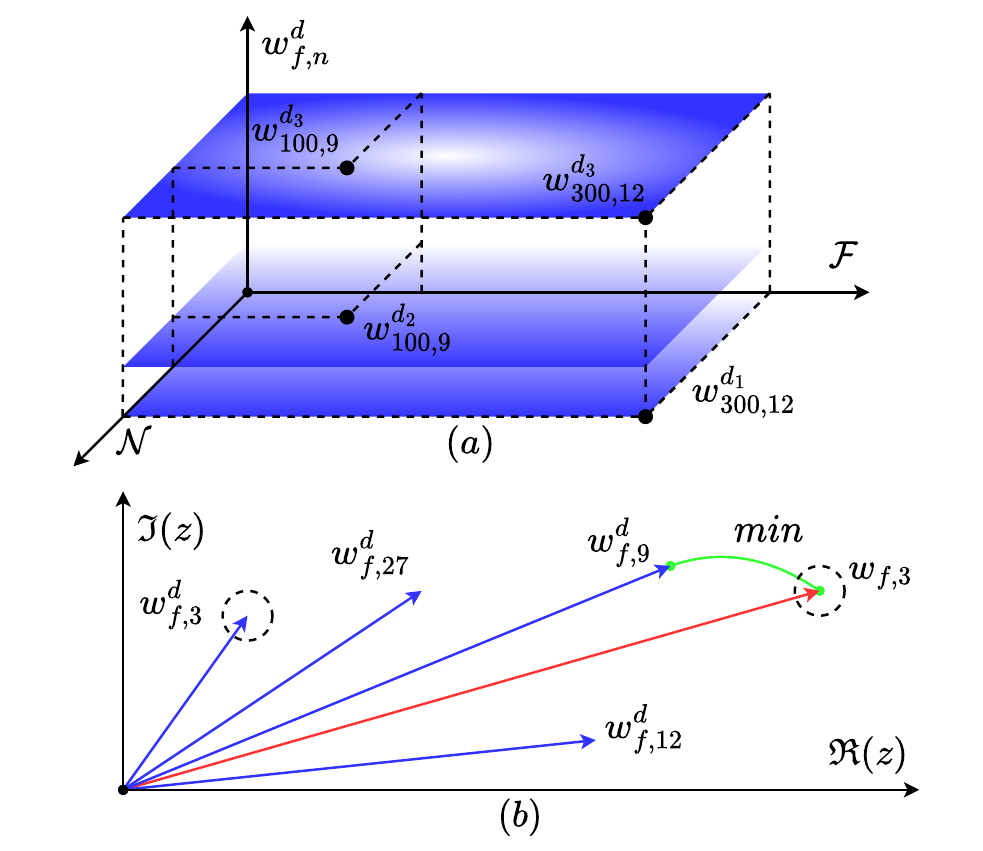}
    \caption{(a) -- Function space, derived from $S_{12}/S_{21}$ measurements. (b) -- Calibration procedure in \gls{abf} systems based on Euclidean distances between $w_{f, n}$ and $w_{f, n}^d$.}
    \label{FIG_calib_space}
\end{figure}

We consider that $S_{12}/S_{21}$ measurements form a function space $f$ (Fig. \ref{FIG_calib_space}\textcolor{blue}{a}), such that $f: \mathcal{F} \times \mathcal{N} \times \mathcal{Z} \rightarrow \mathcal{Y}$, where $\mathcal{F}, \mathcal{N} \in \mathbb{R}$ are domains of continuous and discrete values corresponding to operating frequencies and \gls{rf} channel numbers, respectively.
$\mathcal{Z}, \mathcal{Y} \in \mathbb{C}$ are continuous (for \gls{dbf}) or discrete (for \gls{abf}) domains of undistorted and distorted \gls{bfw}.
Assuming significant sparsity or irregularity of measurements, our problem becomes one of a functional approximation.
In other words, we address a problem of non-linear regression, where given $M$ evaluations $y=\left\{\mathbf{x}_m, f\left(\mathbf{x}_m\right)\right\}_{m=1}^M$, we want to recover the underlying function.
Once recovered, it can be used for \gls{aas} calibration, thus remedying the main disadvantage of conventional ``probe calibration'' methodology.
In addition, three assumptions need to be noted with regard to the regression setting.
First, measurements with corresponding \gls{bfw} are $\mathbb{C}$--valued, so in order to simplify the problem, we solve the regression for $\Re(z)$ and $\Im(z)$ domains, separately. This widens the range of applicable regression models.
Second, by treating $\mathcal{N}$ as a continuous domain, our functional space is assumed to be smooth.
Third, we assume that the underlying \gls{rf} impairments are static, in a sense that measurements do not contain changepoints and the system does not undergo a distribution shift.
Thus, our \textit{\textcolor{red}{third question}} is: \textit{which data-driven model can satisfy the criteria of sample efficiency, robustness and convergence guarantees?}

\section{AAS Calibration via Gaussian Process Models}\label{SOLUTION}
Based on the above considerations, 
our goal is to approximate and interpolate an underlying function in a data-driven fashion.
A wide range of candidate methods includes polynomial regression, splines, reproducing kernel Hilbert spaces and \glspl{dnn}.
However, they only provide deterministic error bounds, and the latter relies on large datasets.
Moreover, our data can be very sparse, which renders deterministic models undesirable.
The alternative that satisfies all our requirements is a \gls{gp} regression which belongs to a class of Bayesian non-parametric models \cite{GaussianProcessesMachine_Rasmussen.Williams_2005}, and represents an unknown function $f(\mathbf{x})$ by a realization sampled from a stochastic process with corresponding stochastic error bounds and convergence guarantees \cite{ConvergenceGaussianProcess_Teckentrup_2020, ConvergenceguaranteesGaussian_Wynne.etal_2022}.
Before proceeding with \gls{aas} calibration solution, we first provide necessary prerequisites for \glspl{gp}. 

\subsection*{\textbf{\textit{Gaussian Process Regression.}}}

\Glspl{gp} define a distribution over functions, such that any finite collection of random variables $f$ has a joint Gaussian distribution (\ref{gpr_principle})

\begin{equation}
    f(\mathbf{x})=\left[f\left(\mathbf{x}_1\right), \ldots, f\left(\mathbf{x}_m\right)\right]^{\top} \sim \mathcal{N}(\boldsymbol{\mu}, k).
    \label{gpr_principle}
\end{equation}

Our dataset $\mathcal{D}$ consists of test input vectors $\mathbf{x}^*$ and noisy samples $\mathbf{y} = \{y_1\left(\mathbf{x}_1\right), \dots, y_m\left(\mathbf{x}_m\right)\}$, corrupted by Gaussian noise $\varepsilon_i \sim \mathcal{N}\left(0, \sigma_{\text {noise }}^2\right), i=1,2, \ldots, m$, that are observed at $\mathbf{x} = \{\mathbf{x}_1, \dots, \mathbf{x}_m\}$, respectively.
Modelling properties of a \gls{gp} are determined by its mean $\boldsymbol{\mu}$ (assumed $\boldsymbol{\mu} = 0$) and covariance kernel function $k\left(\mathbf{x}, \mathbf{x}^{\prime}\right)$.
In the resulting model (\ref{gpr_a}), the prior and measurement models are Gaussian, leading to a Gaussian posterior and a learning problem that amounts to the computation of the conditional mean $\mathbb{E}\left[f\left(\mathbf{x}^*\right)\right]$ and covariances $\mathbb{V}\left[f\left(\mathbf{x}^*\right)\right]$ (\ref{eq_covmean}) of the process at $\mathbf{x}^*$ in closed forms:

\begin{equation}
    f(\mathbf{x}) \sim \mathcal{GP}\left(\boldsymbol{\mu}, k\left(\mathbf{x}, \mathbf{x}^{\prime}\right)\right) \ \ y_i = f\left(\mathbf{x}_i\right)+\varepsilon_i,
    \label{gpr_a}
\end{equation}
\begin{equation}
    \begin{gathered}
    \mathbb{E}\left[f\left(\mathbf{x}^*\right)\right]=\mathbf{k}^{* \top}\left(\mathbf{K}+\sigma_{\text {noise }}^2 \mathbf{I}_m\right)^{-1} \mathbf{y}, \\
    \mathbb{V}\left[f\left(\mathbf{x}^*\right)\right]=k\left(\mathbf{x}^*, \mathbf{x}^*\right)-\mathbf{k}^{* \top}\left(\mathbf{K}+\sigma_{\text {noise }}^2 \mathbf{I}_m\right)^{-1} \mathbf{k}^*,
    \end{gathered}
    \label{eq_covmean}
\end{equation}

where $\mathbf{K}_{i, j}=k\left(\mathbf{x}_i, \mathbf{x}_j\right)$ is $m \times m$ positive semi-definite covariance matrix, $\mathbf{k}^*$ is a $m$--dimensional vector with each entry being $k\left(\mathbf{x}^*, \mathbf{x}_i\right)$.
With (\ref{eq_covmean}), the learning process is formally defined as (\ref{gp_infer}) \cite{GaussianProcessesMachine_Rasmussen.Williams_2005}

\begin{equation}
    p\left(f\left(\mathbf{x}^*\right) \mid \mathcal{D}\right) = \mathcal{N}\left(f\left(\mathbf{x}^*\right) \mid \mathbb{E}\left[f\left(\mathbf{x}^*\right)\right], \mathbb{V}\left[f\left(\mathbf{x}^*\right)\right]\right).
    \label{gp_infer}
\end{equation}

The generalisation and functional properties (e.g. smoothness, periodicity, non-linearity) of \glspl{gp} are encoded by covariance kernels and their hyperparameters $\theta$, which can be optimised by maximizing the marginal likelihood of the data, conditioned on $\theta$ and $\mathbf{x}$.
An example of a commonly used covariance is a \gls{rq} kernel, (\ref{rqk}) where $\theta$ includes length-scale $\ell$, magnitude $\sigma^2$ and $\alpha$ that determines the relative weighting of the large-scale and small-scale variations:

\begin{equation}
    k_{\mathrm{RQ}}\left(\mathbf{x}, \mathbf{x}^{\prime}\right)=\sigma^2\left(1+\left({\left(\mathbf{x}-\mathbf{x}^{\prime}\right)^2}/{2 \alpha \ell^2}\right)\right)^{-\alpha}.
    \label{rqk}
\end{equation}

\subsection*{\textbf{\textit{Kronecker Extension.}}}

In order to support regression over a three-dimensional setting, we apply fast scalable Kronecker inference, originally proposed by Flaxman \etal in \cite{FastKroneckerInference_Flaxman.etal_2015}.
The combination of \glspl{gp} and Kronecker methods relies on the assumption that the kernel in the former is a product of kernels applied across input dimensions allocated on the Cartesian product grid,
without constraints on the grid's regularity and equal cardinality of its axis.
This principle allows to decompose the covariance matrix $\mathbf{K}$ into a Kronecker product $\mathbf{K}=\mathbf{K}_1 \otimes \dots \otimes \mathbf{K}_D$.
Kronecker methods are extended in \cite{FastKroneckerInference_Flaxman.etal_2015} by utilising the Laplace approximation to model the posterior distribution of the \gls{gp} \cite{GaussianProcessesMachine_Rasmussen.Williams_2005}, while also involving the linear conjugate gradients for inference and a lower bound on the \gls{gp} marginal likelihood for kernel learning.
Moreover, their runtime and storage complexities are $\mathcal{O}\left(D m^{\frac{D+1}{D}}\right)$ and $\mathcal{O}\left(D m^{\frac{2}{D}}\right)$, for $D$ input dimensions with $m$ data points, which 
makes them advantageous for efficient learning and implementation.

\subsection*{\textbf{\textit{Calibration Procedure.}}}

We partition the \gls{aas} calibration task into two parts.
The first part consists of creating a sparse measurement dataset $\mathcal{D}$ over a three-dimensional uniform grid $\mathcal{F} \times \mathcal{N} \times \mathcal{Z}$, where each entry $w_{f, n}^d$ is converted from $S_{12}/S_{21}$.
Here we assume that a single measurement captures the full operating bandwidth.
Next, the $\Re(w)$ and $\Im(w)$ components are approximated separately by two \gls{gp} models.
In the second part, the learnt models are deployed at the \gls{aas} and sampled each time the \glspl{bfw} change.
For \gls{dbf} systems, the \gls{bfw} are adjusted in accordance with conventional methods that rely on near-field measurements. 
In the case of \gls{abf}, the situation is different, since the \glspl{bfw} are chosen from a discrete set.
In our scheme, information about the distortion of this set is obtained by surrogate measurements sampled from a \gls{gp} model.
Thus, the task of \gls{abf} calibration is to choose $w_{f, n}^d$ closest to the current $w_{f, n}$ based on Euclidean distances (Fig. \ref{FIG_calib_space}\textcolor{blue}{b}) or, alternatively, to choose the set $w_{f, n}^d$ that best describes the ratios between $w_{f, n}$, such as ${w_{f, 1}}/{w_{f, 2}} \dots {w_{f, n}}/{w_{f, n-1}}$.

\section{Simulation Results}\label{SIMRES}
We model $\Re(w^d)$ and $\Im(w^d)$ components of \gls{rf} impairments as a set of smooth functions $f(\mathcal{F},\mathcal{N})$ defined by a finite Fourier series with random coefficients \cite{SmoothRandomFunctions_Filip.etal_2019}.
Where total number of functions is determined by a range of \gls{bfw} in \gls{aas} and, in case of \gls{abf}, discretisation precision.
In our experiments, \gls{gp} models utilise a Kronecker product of two \gls{rq} and one spectral mixture kernels, with parameters optimised according to \cite{FastKroneckerInference_Flaxman.etal_2015}.

\begin{figure}[]
    \centering
    \captionsetup{justification=centering}
    \includegraphics[scale=0.33]{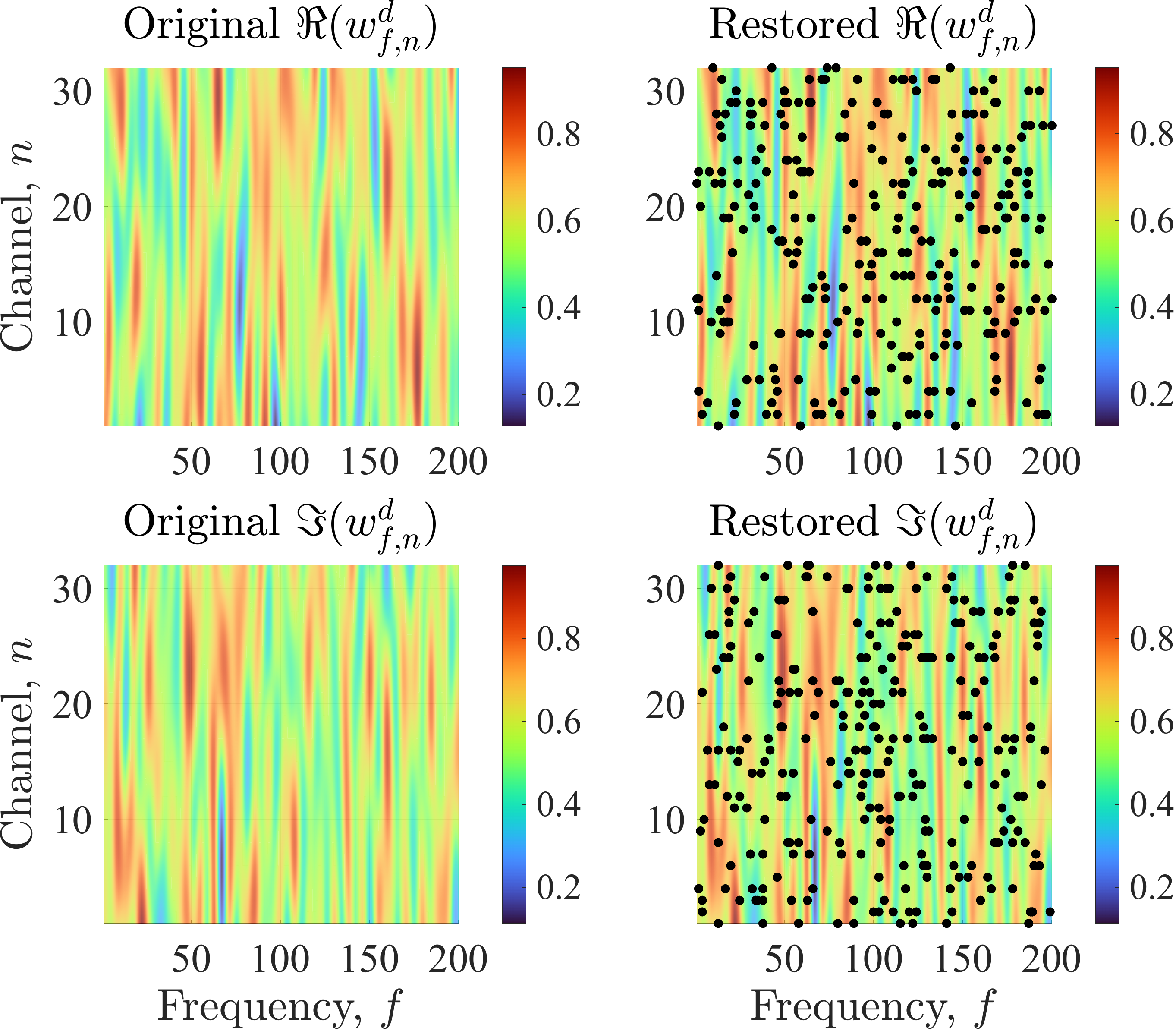}
    \caption{Slices of $w^d$ space for 32 $\times$ 32 antenna array and 400 MHz operating bandwidth. Colours indicate magnitude and black dots -- locations of observed samples for \gls{gp} modelling.}
    \label{gp_surf_test}
\end{figure}

As an example, in Fig. \ref{gp_surf_test} we show impairment surfaces and their \gls{gp} approximations, for a \gls{abf} with $\mathcal{N}=1024$, $\mathcal{F}=400$ MHz (with granularity of $1$ MHz) and 1024 \gls{bfw} $w_{f, n}$.
For the purpose of visual clarity, we show a ``slice'' of impairment space, corresponding to a single $w_{f,n}$ out of $1024$.
Left half of subplots shows ground-truth surfaces of $\Re(w_{f, n}^d)$ and $\Im(w_{f, n}^d)$ values. 
While right half -- results of \gls{gp} modelling based on randomly sampled $w_{f,n}^d$ observations.
We compare the original and restored surfaces according to the normalised \gls{rmse} criterion averaged across a series of $5 \times 10^4$ random $w_{f, n}^d$ space-realisations, in which models utilising $5, 10, 15 \text{ and } 20\%$ of all possible measurements of $w_{f,n}^d$ were compared.
With proposed approach best approximation quality was achieved for $15\%$ and $20\%$ utilisation, with an averaged \gls{rmse} of $2.3710e^{-04}$ and $1.0380e^{-04}$ respectively.

After restoring $w_{f,n}^d$ with the \gls{gp} models, we study the performance of \gls{aas} calibration in both \gls{abf} and \gls{dbf} at different scales.
In all cases, the \glspl{aas} use rectangular arrays with uniform element spacing of $0.5 \lambda$ and \glspl{gp} with an approximation \gls{rmse} lower than $10^{-3}$. 
All beam patterns, depicted below, show the azimuth $\phi^{\circ}$ and elevation $\theta^{\circ}$ cuts in the center of the operating bandwidth of 400 MHz and share the same notation.
The desired beam patterns (based on $w_{f,n}$) are depicted by black curves, the distorted ones (based on $w_{f,n}^d$) by blue curves and the calibrated ones (based on knowledge of the restored $w_{f,n}^d$) by dashed red curves.
The \Gls{bfw} are synthesised using convex optimisation with constraints on angular positions of \gls{ue} (magenta dots) and interference sources (black dots).

\begin{figure}[]
    \centering
    \includegraphics[scale=0.35, keepaspectratio]{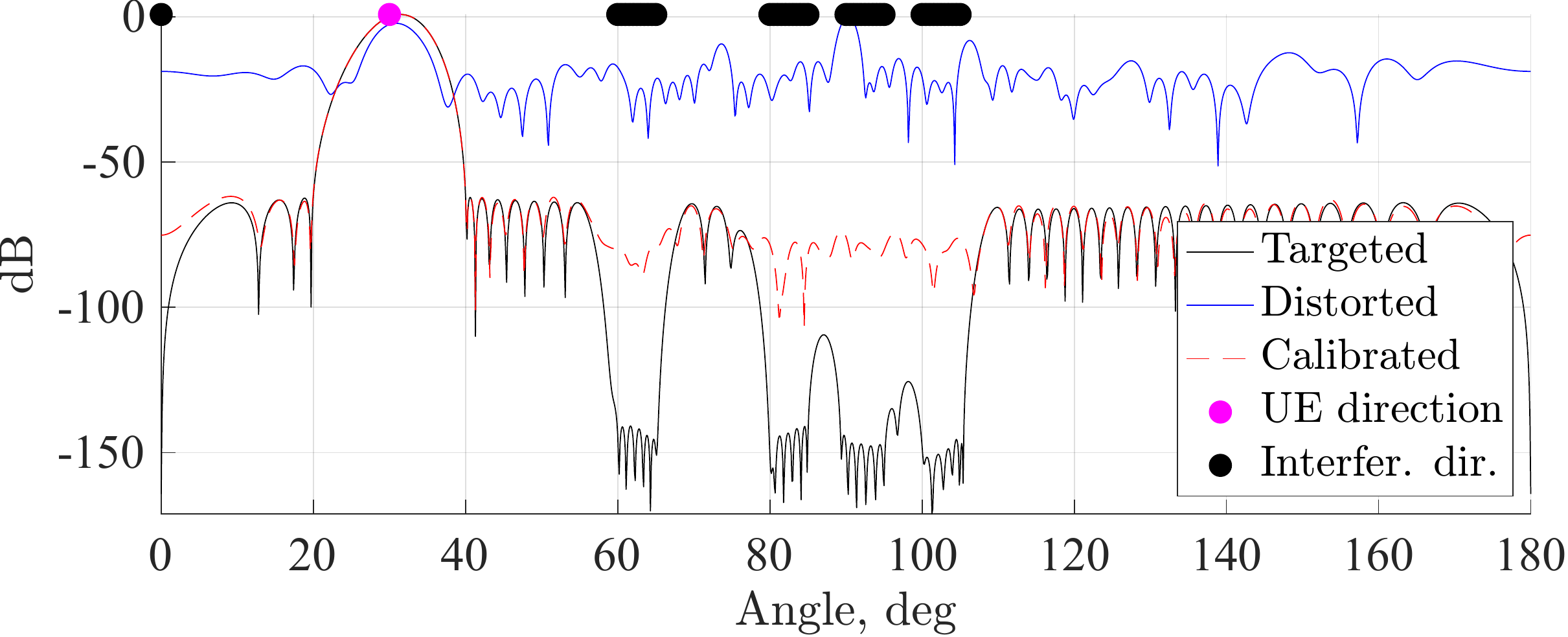}
    \includegraphics[scale=0.35, keepaspectratio]{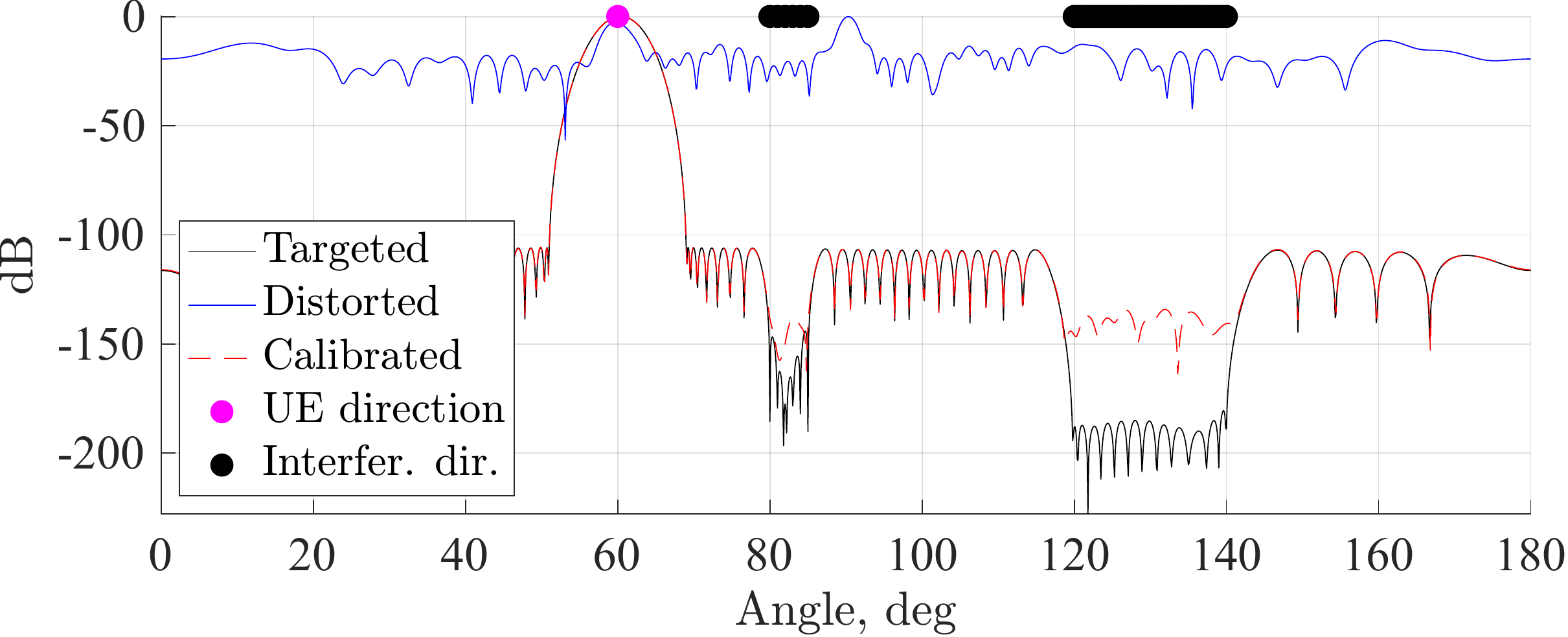}
    \caption{
        \Gls{abf} with $64 \times 64$ element array; 
        \gls{ue} $\phi^{\circ},\theta^{\circ}=(30; 60)$;
        interference sets: 
        $\phi^{\circ}= [60^{\circ},65^{\circ}], ..., [100^{\circ},105^{\circ}]$ and $\theta^{\circ}=[80^{\circ},85^{\circ}], [120^{\circ},140^{\circ}]$.
    }
    \label{abf_b}
\end{figure}

\begin{figure}[]
    \centering
    \includegraphics[scale=0.35, keepaspectratio]{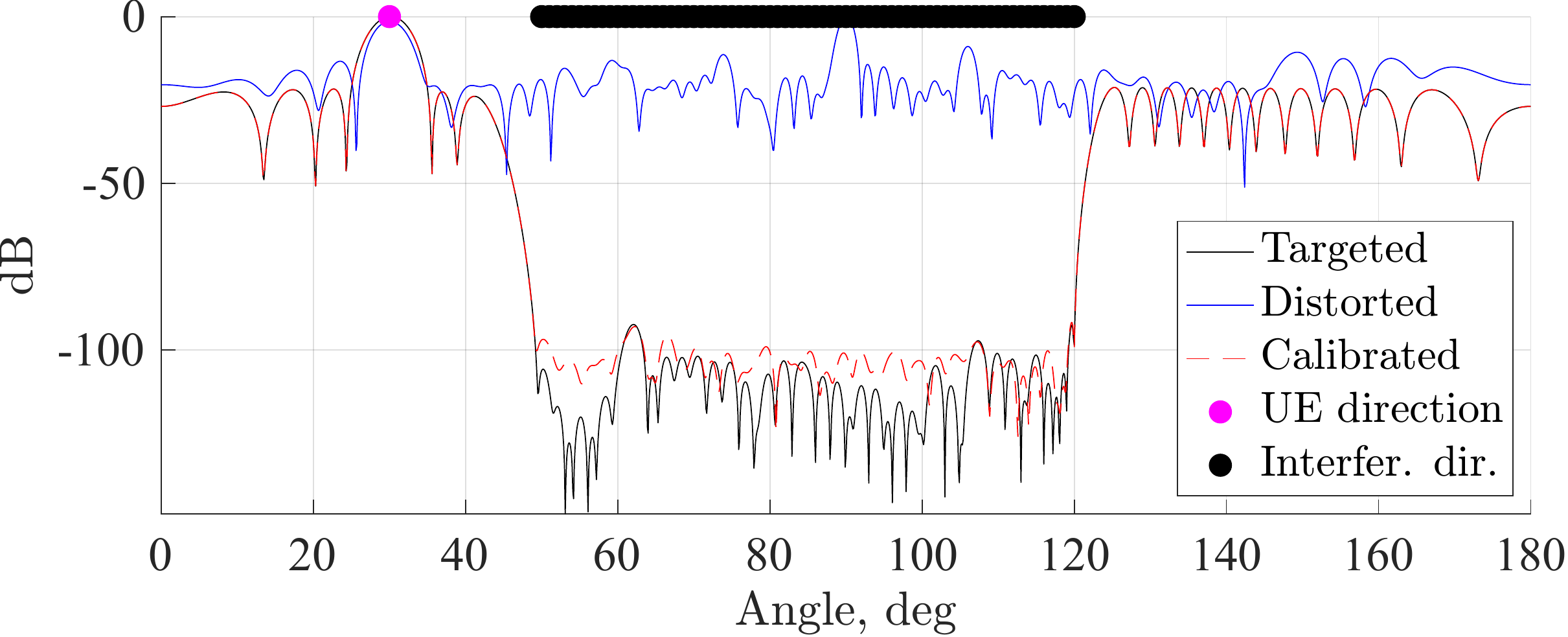}
    \includegraphics[scale=0.35, keepaspectratio]{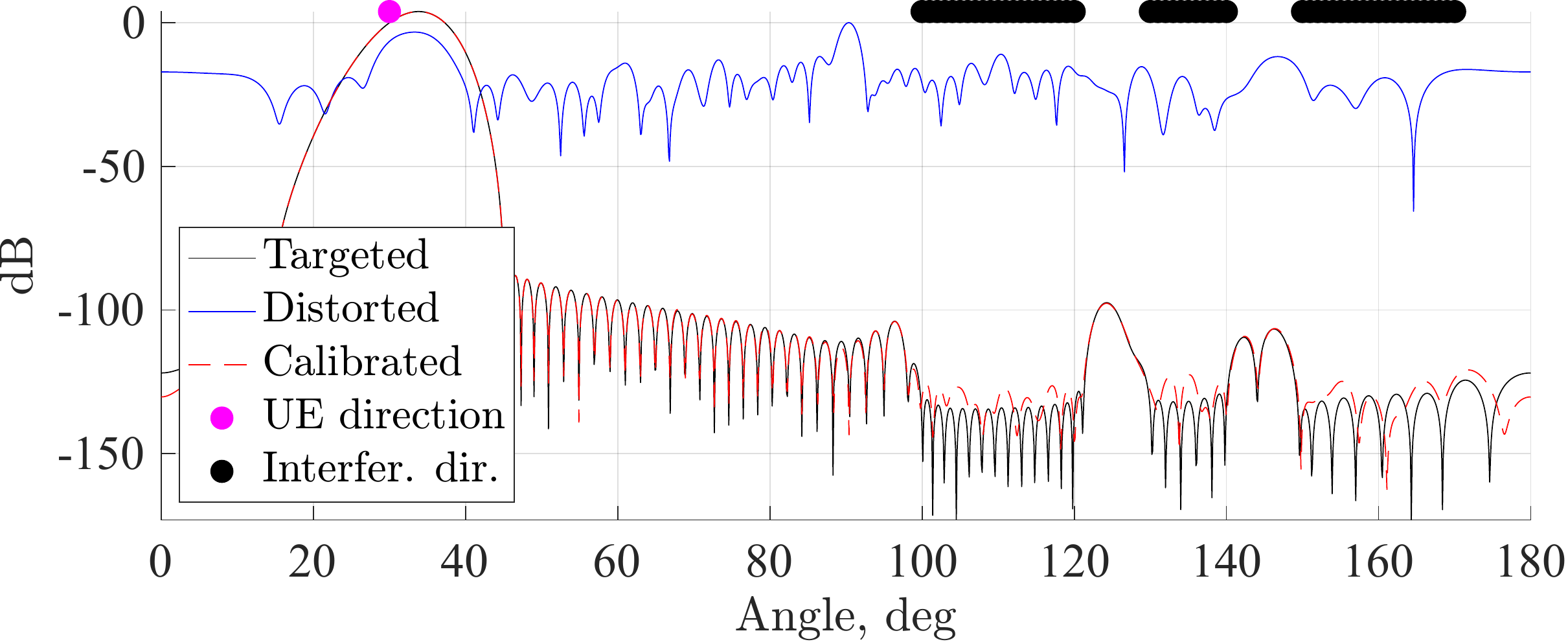}
    \caption{
        \Gls{dbf} with $64 \times 64$ element array;
        \gls{ue} $\phi^{\circ},\theta^{\circ}=(30; 30)$;
        interference sets: $\phi^{\circ}=[50^{\circ},120^{\circ}]$ and $\theta^{\circ}=[100^{\circ},120^{\circ}], ..., [150^{\circ},170^{\circ}]$.
    }
    \label{dbf_a}
\end{figure}

\begin{figure}[]
    \centering
    \includegraphics[scale=0.35, keepaspectratio]{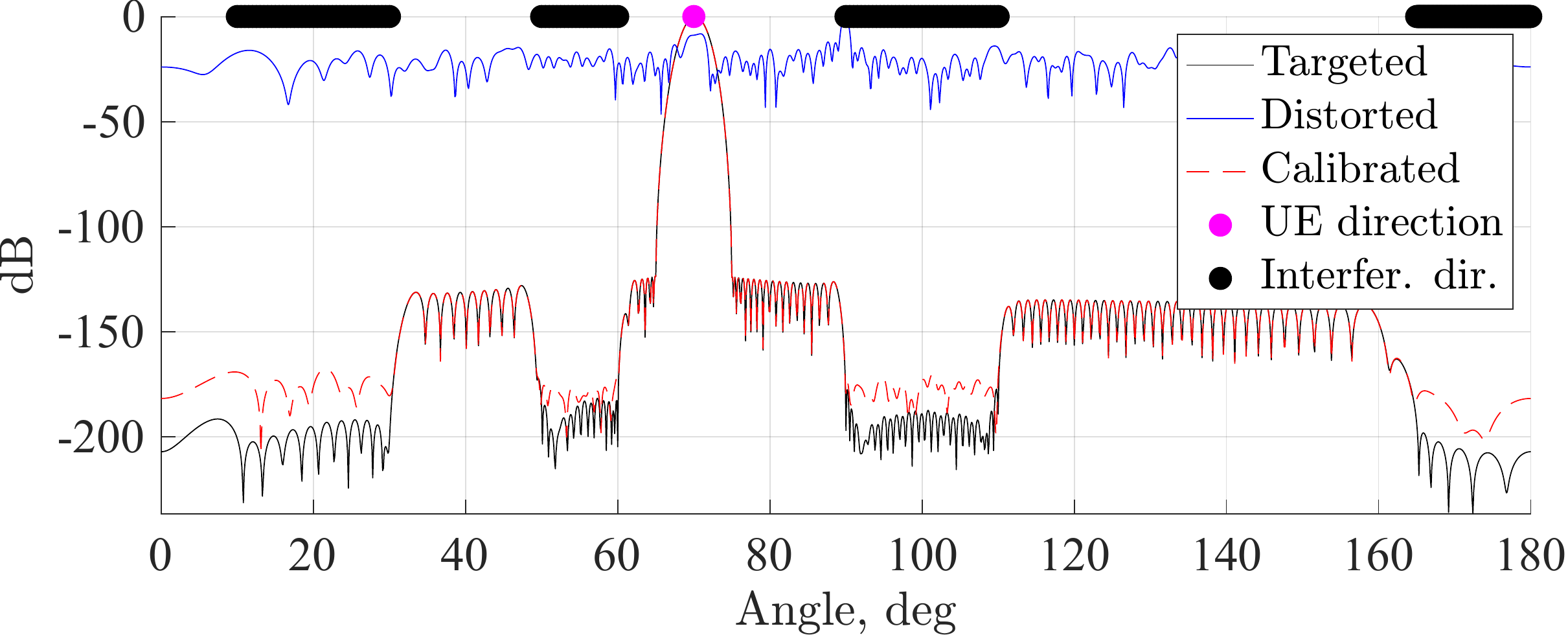}
    \includegraphics[scale=0.35, keepaspectratio]{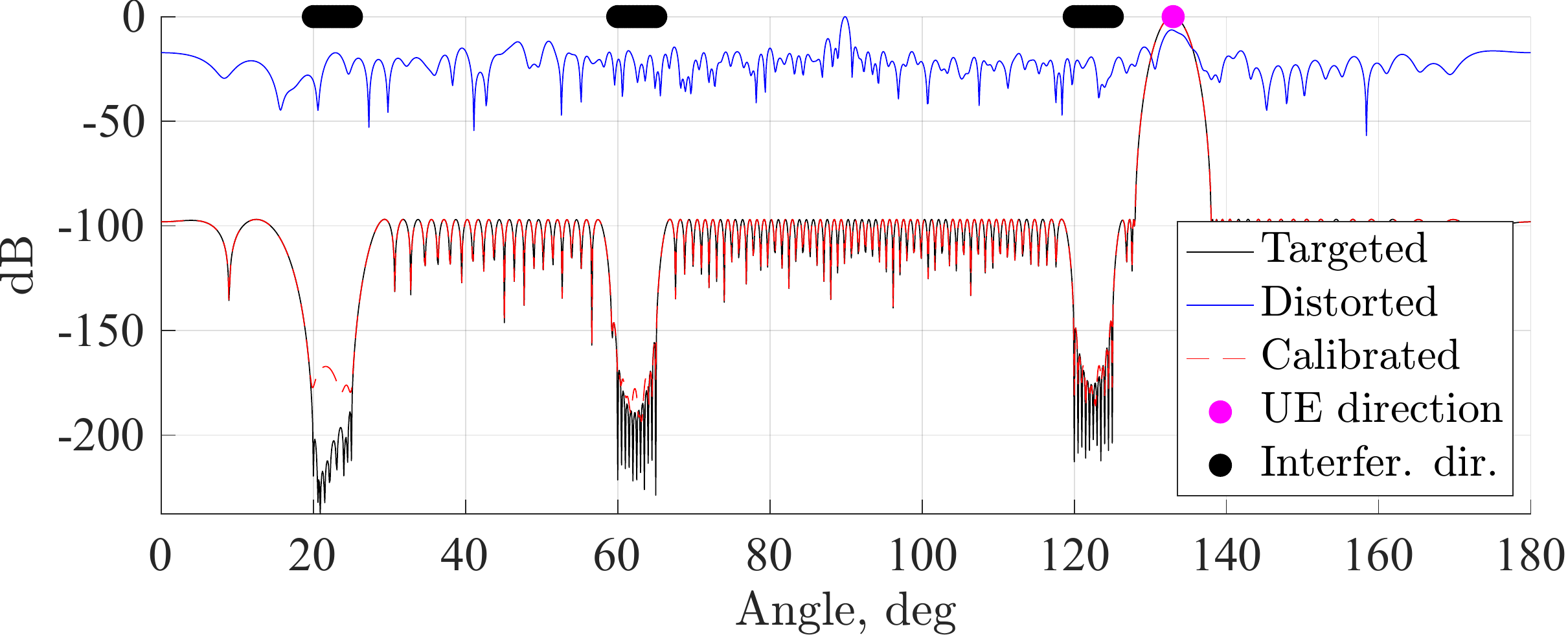}
    \caption{
        \Gls{dbf} with $128\times128$ element array; 
        \gls{ue} $\phi^{\circ},\theta^{\circ}=(70; 133)$;
        interference sets $\phi^{\circ}=[10^{\circ},30^{\circ}], ..., [165^{\circ},180^{\circ}]$ and $\theta^{\circ}=[20^{\circ},25^{\circ}], ..., [120^{\circ},125^{\circ}]$.
    }
    \label{dbf_b}
\end{figure}

Continuing the \gls{abf} case with 1024 $w_{f, n}$, we show calibration performance for \gls{abf} systems with 1024 and 4096 antenna elements, with the latter shown in Fig. \ref{abf_b}. 
In both scenarios the \gls{bpa} \gls{rmse} improvement ratio was significantly hindered by unavoidable discretisation of \gls{bfw}, resulting only in $BPA^{ABF} / BPA_{calib.}^{ABF} = 2.23$ and $1.91$ respectively.
Lastly, we switch to \gls{dbf} arrays with $64 \times 64$ (Fig. \ref{dbf_a}) and $128 \times 128$ (Fig. \ref{dbf_b}) elements, where in the same experiment setting $BPA^{DBF} / BPA_{calib.}^{DBF}$ achieves values of 
$8.11$ and $10.04$.

\section{Conclusions and Future Work}\label{FIN}
In this work, we have proposed a novel approach to \gls{aas} calibration based on the intersection of near-field measurement methodologies and Bayesian modelling.
Our solution captures the complete set of internal and external \gls{rf} hardware impairments via a sparse set of precise near-field $S_{12}/S_{21}$ parameter measurements and restores the rest via high-dimensional \gls{gp} regression models.
Once deployed at an operating \gls{aas}, the latter actively adjusts the beamforming parameters in accordance with the learnt impairment values and thus results in corrected beam patterns and a consequential performance improvement of algorithms that rely on their high fidelity.
Furthermore, it avoids the drawbacks of conventional methods, such as relying on \gls{rf} hardware constraints or standardised communication protocols and signal formats.

Based on the obtained results and the modelling properties of \glspl{gp}, we envision a future work in the context of three key directions: extending the solution to \gls{hbf} architectures, modelling of \gls{rf} impairments with temporal dynamics related to temperature drifts or hardware degradation and lastly, incorporating uncertainty quantification into the calibration process.
Finally, as we continue the line of thought about algorithm design and evaluation, we argue that the domain of \gls{aas} calibration would benefit from a unified benchmarking test and performance metric that will allow researchers to analyse the properties of their proposals in an unbiased and consistent way.

\section*{Acknowledgements}
The work of S. S. Tambovskiy is funded by the Marie Skłodowska Curie action WINDMILL (grant No. 813999). G. Fodor was supported by the Swedish Strategic Research (SSF) grant for the FUS21-0004 SAICOM project.

\balance
\bibliography{sergeyta_wsa23_conf}

\end{document}